\documentclass{icsm}

\slacs{.4ex}

\begin{document}
\def\lie{{\cal G}}
\def\v{\vert}
\def\bv{\bigm\vert}
\def\nonu{\nonumber}
\def\h{ {1\over 2}  }
\def\eps{\epsilon}
\def\a{\alpha}
\def\b{\beta}
\def\d{\delta}
\def\p{\phi}
\def\l{\lambda}
\def\pr{\prime}
\def\br{\begin{eqnarray}}
\def\er{\end{eqnarray}}
\def\pa{\partial}
\def\o{\over}
\def\Ker{\mathop{\mathrm{Ker}}}
\def\ad{\mathop{\mathrm{ad}}}
\def\Tr{\mathop{\mathrm{Tr}}}

\title{Algebraic construction of integrable and super integrable hierarchies}
\authori{H. Aratyn}
\addressi{Department of Physics, University of Illinois at Chicago\\
845 W. Taylor St., Chicago, Illinois, 60607--7059}
\authorii{J.F. Gomes and A.H. Zimerman}
\addressii{Instituto de F\'\i sica Te\'orica -- IFT/UNESP\\
Rua Pamplona 145, 01405--900, S\~ao Paulo - SP, Brazil}
\authoriii{}    \addressiii{}
\authoriv{}     \addressiv{}
\authorv{}      \addressv{}
\authorvi{}     \addressvi{}
%
%Page headings:
\headauthor{H. Aratyn, J.F. Gomes and A.H. Zimerman}
\headtitle{Algebraic construction of integrable and super
integrable hierarchies}
\lastevenhead{H. Aratyn, J.F. Gomes and A.H. Zimerman: Algebraic
construction of integrable \ldots}
\maketitle

\begin{abstract}
A general construction of integrable hierarchies based on affine
Lie algebras is presented. The models are specified according to
some algebraic data and their time evolution is obtained from
solutions of the zero curvature condition. Such framework provides
an unified treatment of relativistic and non relativistic models.
The extension to the construction of supersymmetric integrable
hierarchies is proposed. An explicit example of $N=2$ super mKdV
and sinh--Gordon is presented.
\end{abstract}

\section{Introduction}

Integrable models consist of a very peculiar class of physical
systems described by non-linear differential equations.  In two
dimensional space time they present an infinite number of
conservation laws and admit soliton solutions. Well known examples
such as the sine--Gordon or mKdV equations have been shown to be
connected to many applications in several branches of physics.

Integrable hierarchies can be constructed and classified in terms
of an affine Lie algebraic structure. Equations of motion for
relativistic integrable models can be derived from the
Leznov--Saveliev's approach \cite{ls}, which  was  latter
discovered to be connected to reductions of the
Wess--Zumino--Witten model \cite{ora}.  For non relativistic
integrable hierarchies, the Lax formulation proved to be an
important framework which also relies on the algebraic
decomposition of an affine Lie algebra.

The aim of this paper is to discuss a universal formulation which embeds
both relativistic and non relativistic  integrable models can be
viewed. The main ingredient is  the zero curvature condition which
employs, in a natural manner,  Lie algebra valued functionals of
the physical fields.

We first discuss  the Leznov--Saveliev construction of
relativistic integrable models in terms of a decomposition of an
affine Lie algebra. The models are shown to be classified
according to  some algebraic data, namely
$(Q,\eps_{\pm},\lie_0^0)$, where $Q$ specifies the space of
physical fields, $\eps_{\pm}$, the interaction and
$\lie_0^0=\mathcal{K}=\Ker(\ad_{\eps_+})$ contains the internal
soliton symmetries. Next, we discuss the Lax formulation  for the
non-relativistic integrable models in terms of certain positive
grade elements in ${\cal K}$, each of them is assigned to a time
evolution. The precise correspondence between those two
formulations is explained in Sect. 3 where  we introduce negative
grade time evolution and show that the relativistic formulation of
Leznov and Saveliev corresponds to the  grade $-1$ generator in
${\cal K}$.  In Sect. 4 we extend both formulations to accommodate
affine decompositions with integer and half integer gradings. This
leads to the construction of supersymmetric integrable
hierarchies. Finally as an example, we  construct  the $N=2$ super
MKdV and $N=2$ super sinh--Gordon models associated to
$\widehat{SL}(2,2)$ showing  that they indeed belong to the same
integrable hierarchy.

\section{General construction of relativistic integrable hierarchies}

A general construction of relativistic integrable hierarchies in
terms  of an affine  Lie algebra $\hat{\lie} $ can be established
from the Leznov--Saveliev equations of motion \cite{ls}, \be
\bar\pa (B^{-1}\pa B)+[\eps_-, B^{-1}\eps_+B]=0\,,\quad
\pa(\bar\pa B B^{-1})-[\eps_+, B\eps_-B^{-1}]=0\,, \label{1.1} \ee
where the space time is represented by of the light cone
coordinate $z=t+x$, $\bar z=t-x$ and a decomposition of
$\hat{\lie}$ into graded subspaces $\hat{\lie}=\oplus_a{\lie_a}$,
$a\in Z$ according to a grading operator $Q$ such that
$[Q,{\lie_a}]=a{\lie_a}$ is assumed. Here $B$ represents an
element of the zero grade subgroup $B \in G_0$ and is parametrized
by the physical fields (Toda fields) of the theory.  $\eps_{\pm}$
are constant generators of grade $\pm 1$  which characterize the
non linear interaction. The classical integrability of  these
models follows from their zero curvature (Lax) representation: \be
\pa\bar{{\cal A}}-\bar\pa{\cal A}+[{\cal A},\bar {{\cal A}}]=0\,,
\quad {\cal A},\bar{{\cal A}}\in\oplus_{i=0,\pm 1}\lie_i\,,
\label{zzc} \ee with \be {\cal A}=B\eps_-B^{-1}\,,\quad
\bar{{\cal A}}=-\eps_+-\bar\pa B B^{-1}\,. \label{zz} \ee
The existence of an infinite set (of commuting) conserved charges
$P_m$, $m=0,1,\cdots$ is a direct consequence of eqn. (\ref{zzc}),
namely,
$$
P_m(t)=\Tr\bigl(T(t)\bigr)^m\,,\quad \pa_tP_m=0\,,\quad
T(t)=\lim_{L\to\infty}{\cal P} \exp\biggl(\int_{-L}^{L}{\cal
A}_x(t,x)\D x\biggr),
$$
where ${\cal A}_x={{\cal A}}+\bar{{\cal A}}$. Well known examples
of integrable models fall into the general construction as we
shall now detail.

Consider $\hat{\lie}=\widehat{SL}(2)$ with the principal
gradation, i.e. $Q=2\D+{1\o 2}H$ and $\D$ is the derivation
operator. The zero grade subalgebra is then $\lie_0=\{ H^{(0)}\}$
and hence $B=\E^{\phi H^{(0)}}$. If we now choose
$\eps_+=E_\a^{(0)}+ E_{-\a}^{(1)}$, $\eps_-=\eps_+^{\dagger}$ we
find from (\ref{zzc}) the sinh--Gordon equation \be
\pa\bar{\mathcal{A}}-\bar\pa {\cal A} + [{\cal A},\bar {{\cal A}}]
=\left(\bar\pa\pa\phi+\E^{2\phi}-\E^{-2\phi}\right)H=0\,.
\label{1.2} \ee Another decomposition of $\widehat {SL}(2)$ can be
obtained from the homogeneous  gradation where $Q=\D$. The zero
grade subalgebra acquires  a nonabelian structure, i.e.
$\lie_0=\{H^{(0)},E_{\pm \a}^{(0)}\}$ and
$B=\E^{RH^{(0)}/2}\E^{\chi E_{-\a}^{(0)}}\E^{\psi E_{\a}^{(0)}}
\E^{R H^{(0)}/2}$. If we now choose $\eps_{\pm}=H^{(\pm)}$ we find
from eqns. (\ref{1.1}) the existence of two chiral currents
associated to the subalgebra $\lie_0^0\subset\lie_0$ namely,
$\bar\pa J_{H}=\bar\pa\Tr(H^{(0)}B^{-1}\pa B)=0$ and $\pa\bar
J_{H}=\pa\Tr(H^{(0)}\bar\pa BB^{-1})=0$, since
$\lie_0^0=U(1)=\{H^{(0)}\}$, $[\lie_0^0,\eps_{\pm}]=0$. This fact
allows the implementation of the following subsidiary constraints
\be J_{H}=\pa R-\frac{\psi\pa\chi}{\Delta}=0\,,\quad \bar J_{H}
=\bar\pa R-\frac{\chi\bar\pa\psi}{\Delta}=0\,,\quad
\Delta=1+\psi\chi\,, \label{1.3} \ee which eliminates the nonlocal
field $R$. The equations of motion for this case are then given by
the zero curvature condition (\ref{zzc}) or, equivalently by eqns.
(\ref{1.1}), when the subsidiary constraints (\ref{1.3}) are taken
into consideration (i.e. for the coset $G_0/G_0^0 = SL(2)/U(1)$),
yielding the equations of motion of Lund--Regge (complex
sine--Gordon) model, \be
\bar\pa\left(\frac{\pa\chi}{\Delta}\right)+
\frac{\chi\pa\chi\bar\pa\psi}{\Delta^2}+2\chi=0\,,\quad
\pa\left(\frac{\bar\pa\psi}{\Delta}\right)+
\frac{\psi\pa\chi\bar\pa\psi}{\Delta^2}+2\psi=0\,.\label{1.4} \ee
The gauged $\lie_0^0 = U(1)$ factor arisen from the constraints
(\ref{1.3}) is responsible for the global $U(1)$ symmetry
$\psi\rightarrow\psi\E^{\a}$, $\chi\rightarrow\chi\E^{-\a}$ and
henceforth for the conservation of the electric charge \be
Q_{\mathrm{el}}=\int_{-\infty}^{+\infty}(\pa_x R)\,\D x\,.
\label{1.5} \ee

The above examples can be extended to higher rank affine algebras
with gradations which are intermediate between the principal and
homogeneous ones. Let us consider $\lie = \widehat {SL}(n+1)$. In
general, we can construct integrable hierarchies according to the
following algebraic structures:
\begin{enumerate}
\item
$\lie_0^0=\emptyset$ characterizes the choices of \br
Q&=&(N+1)\D+ \sum_{l=1}^{N} \l_l\cdot H\,,\quad
\lie_0 = U(1)^N = \{h_1, \cdots , h_N \}\,, \nonu \\
\eps_{\pm}&=&\sum_{l=1}^{N}E_{\pm \a_l}^{(0)}+
E_{\mp(\a_1+\cdots+\a_N)}^{(\pm 1)}\,, \nonu \er which gives rise
to the well known abelian affine Toda model (see for instance
\cite{ora}, \cite{ls}).
\item
\begin{enumerate}
\item
$\lie_0^0=U(1)=\{\l_1\cdot H\}$ \br
Q&=&N\D+\sum_{l=2}^{N}\l_l\cdot H\,,\quad
\lie_0=SL(2)\otimes U(1)^{N-1}=\{E_{\pm \a_1},h_1,\cdots,h_N\}\,, \nonu \\
\eps_{\pm}&=&\sum_{l=2}^{N} E_{\pm\a_l}^{(0)}+
E_{\mp(\a_2+\cdots+\a_N)}^{(\pm 1)}\nonu \er corresponds to the
simplest non abelian affine Toda model of dyonic type, admitting
electrically charged topological solitons (see for instance
\cite{tau}).
\item $\lie_0^0 =U(1)\otimes U(1)=\{\l_1\cdot H,\l_N\cdot H\}$ \br
Q&=&(n-1)\D+ \sum_{l=2}^{N-1} \l_l\cdot H\,,\quad
\eps_{\pm}=\sum_{l=2}^{N-1}E_{\pm \a_l}^{(0)}+
E_{\mp(\a_2+ \cdots + \a_{N-1})}^{(\pm 1)}\,, \nonu \\
\lie_0&=&SL(2)\otimes SL(2)\otimes U(1)^{N-2}=
\{E_{\pm\a_1},E_{\pm \a_N}, h_1, \cdots , h_N \}\,, \nonu \er is
of the same class of $U(1)^{\otimes k}$ dyonic type IM's, but now
yielding multicharged solitons (\cite{multi}).
\end{enumerate}
\item $\lie_0^0=SL(2)\otimes U(1)=\{ E_{\pm\a_1},\l_1\cdot
H,\l_2\cdot H\}$ \br Q&=&(N-1)\D+\sum_{l=3}^{N}\l_l\cdot H\,,\quad
\eps_{\pm}=\sum_{l=3}^{N}E_{\pm \a_l}^{(0)}+
E_{\mp(\a_3+\cdots+\a_{N})}^{(\pm 1)}\,, \nonu \\
\lie_0&=&SL(3)\otimes
U(1)^{N-2}=\{E_{\pm\a_1},E_{\pm\a_2},E_{\pm(\a_1+\a_2)},
h_1,\cdots,h_N\}\,, \nonu \er leads to dyonic models with non
abelian global symmetries (see \cite{spin}).
\end{enumerate}
In fact the integrable hierarchies are classified in terms of the
gradation $Q$, the constant operators $\eps_{\pm}$ and by the
global symmetry group described by the subalgebra $\lie_0^0 = Ker (ad_{E})={\cal K}$.

\section{Non relativistic construction of integrable hierarchies}

We will follow the approach  given in ref. \cite{kluwer} which
associates to every {\it positive grade} $n$ element $E^{(n)} \in
C({\cal K})$ ($ C ({\cal K}) = \{ x  \in {\cal K} ,  [x, y]=0  \;\; {\rm for \; all} \;\; y \;\; \in {\cal K } \}$)  a
time evolution $t_n$ \be \pa_{t_n} \Theta (t)= (\Theta E^{(n)}
\Theta^{-1})_{-} \Theta (t)\,, \label{3.2} \ee for the dressing
matrix $\Theta=\exp\bigl(\sum_{i<0}\theta^{(i)}\bigr)$ being an
exponential in $\lie_{<}$ and $(\;\;)_{-}$ represents the
projection on strictly negative grades. By construction such flows
commute, i.e. $[\pa_{t_m},\pa_{t_n}]\Theta(t)=0$. In particular
for $n=1$, $\pa_{t_1}\equiv\pa_x$, $\eps_+=E$ we find \br
\pa_x(\Theta)&=&(\Theta E \Theta^{-1})_{-} \Theta=\big\lbrack
\Theta E\Theta^{-1}-(\Theta E\Theta^{-1})_{+}\big\rbrack\Theta= \nonu \\
&=&\Theta E-\left(E+\bigl[\theta^{(-1)},E\bigr]\right)\Theta= \nonu \\
&=&\Theta E-(E+A_0)\Theta\,, \label{3.3} \er where $( \;\; )_{+}$
represents the projection on positive and zero grades and
$A_0=\left[\theta^{(-1)},E\right] $. Clearly $A_0$ is in
$\mathcal{M}$ (the Image of the adjoint operation $ad(E)X = [E,X]$)
 and has  grade zero. This leads to the dressing
expression \be \Theta^{-1}(\pa_x+E+A_0)\Theta=\pa_x+E\label{3.4}
\ee for the Lax operator $L=\pa_x+E+A_0$. Similarly, for higher
flows we obtain \be \Theta^{-1}\biggl(\pa_{t_n}+ E^{(n)}+
\sum_{i=0}^{n-1}D^{(i)}_n\biggr)\Theta=\pa_{t_n}+ E^{(n)}\,,
\label{3.5} \ee where
$$
(\Theta E^{(n)}\Theta^{-1})_+=E^{(n)}+\sum_{i=0}^{n-1}D^{(i)}_n\,.
$$
These dressing relations give rise to the zero--curvature
conditions \be
\biggl[\pa_x+E+A_0,\pa_{t_n}+E^{(n)}+\sum_{i=0}^{n-1}D_n^{(i)}\biggr]=
\Theta \left[\pa_x+E,\pa_{t_n}+E^{(n)}\right]\Theta^{-1}=0\,,
\label{3.6} \ee where
$D^{(j)}_n=D^{(j)}_{n\mathcal{K}}+D^{(j)}_{n\mathcal{M}}\in\lie_j$.
We therefore find \be\label{2.3}\ba{l}
[E,D^{(n-1)}_n]+[A_0,E^{(n)}]=0\,,\\[2pt]
[E,D^{(n-2)}_n]+[A_0,D^{(n-1)}_n]+\pa_x D^{(n-1)}_n=0\,,\\[2pt]
\qquad\vdots\\[2pt]
[A_0,D^{(0)}_n]-\pa_{t_n}A_0+\pa_x D^{(0)}_n=0\,.\ea\ee Each
equation can be decomposed into ${\cal K}$ and ${\cal M}$
components. It is clear that a local solution for $D^{(i)}_n$,
$i=0,\cdots,n$ can be found recursively starting from the highest
grade eqn. in (\ref{2.3})  until we reach the last. In particular
the eqn. corresponding to the zero grade component also gives rise
to the time evolution of the physical fields.

Let us reconsider the examples given in the previous section in
connection with  ${\widehat \lie} = \widehat {SL}(2)$. With $Q $
given in the principal gradation, $Q=2\D+ {1\o 2} H$ and
$\eps_+=E_\a^{(0)}+E_{-\a}^{(1)}$ we parametrize $A_0=u H^{(0)}\in
{\cal M}$ and solve eqn. (\ref{zzc}) for $t=t_3$. After solving
for $D^{(3)}_3$, $D^{(2)}_3$, $D^{(1)}_3$ and $D^{(0)}_3$ we
obtain the equation of motion for the {\it mKdV} model, \be
\pa_{t_3} u = u_{xxx} + 6 u^2 u_x\,. \label{2.4} \ee The
decomposition of ${\widehat \lie} = \widehat {SL}(2)$  according
to the homogeneous gradation $Q=\D$ leads to $A_0=q
E_a^{(0)}+rE_{-a}^{(0)}\in{\cal M}$ and eqn. (\ref{zzc}) yields
for $t=t_2$ the {\it nonlinear Schroedinger} equation (NLS),
\be\label{2.5}\ba{rcl}
\pa_{t_2}q+q_{xx}-2rq^2&=&0\,,\\[2pt]
\pa_{t_2}r+r_{xx}+2qr^2&=&0\,.\ea\ee  Notice that, the
sinh--Gordon and the mKdV equations as well as,  the Lund--Regge
and the NLS equations, are constructed from the same algebraic
structure, i.e., same ${\widehat \lie}= \widehat {SL}(2)$, $Q$ and
same choice of $ E=\eps_+$. This indicates that they belong to the
same integrable hierarchy.  In fact we shall prove such statement
in a more precise manner, but before that, let us discuss the
positive hierarchies associated to the
${\widehat\lie}=\widehat{SL}(n+1)$  generalized models of the
previous section. To each algebraic structure 1. to 3. we
associate the following Lax operators:
\begin{enumerate}
\item $\lie_0^0={\cal K}=\emptyset$
$$
A_0=\sum_{i=1}^{n} u_ih_i^{(0)}\,,
$$
\item
\begin{enumerate}
\item ${\cal K}=U(1)=\{\l_1\cdot H^{(0)}\}$
$$
A_0=qE_{\a_1}^{(0)}+rE_{-\a_1}^{(0)}+\sum_{i=2}^{n}u_ih_i^{(0)}\,,
$$
\item ${\cal K}=U(1)\otimes U(1)=\{\l_1\cdot H^{(0)},\l_n\cdot
H^{(0)}\}$
$$
A_0=q_1E_{\a_1}^{(0)}+q_nE_{\a_n}^{(0)}+r_1E_{-\a_1}^{(0)}+
r_nE_{-\a_n}^{(0)}+\sum_{i=2}^{n-1}u_i h_i^{(0)}\,,
$$
\end{enumerate}
\item ${\cal K}=SL(2)\otimes U(1)=\{E_{\pm\a_1},\l_1\cdot H,\l_2
\cdot H\}$ \be A_0=q_1E_{\a_1+\a_2}^{(0)}+q_2E_{\a_2}^{(0)} +
r_1E_{-\a_1-\a_2}^{(0)}+r_2E_{-\a_2}^{(0)}+\sum_{i=3}^{n}u_ih_i^{(0)}\,,
\ee
\end{enumerate}
they give rise to the generalized mKdV,  multicomponent  AKNS and constrained KP (cKP)
hierarchies \cite{fordy-kulish}, \cite{jmpa}, \cite{jmpb} respectively.

\section{Negative grade time evolution}

The time evolution associated to {\it negative grade} elements in
$ C( {\cal K})$ can be incorporated within the general
construction  of integrable hierarchies following the
Riemann--Hilbert problem  and its connection with the dressing
formulation \cite{jgp}.  As in eqn. (\ref{3.2}) to each element
$E^{(-n)} \in {\cal K}$  we define the associated time evolution
by \be {{\pa \Theta (t)}\o {\pa {t_{-n}}}} = - ( BM
E^{(-n)}M^{-1}B^{-1})_{-} \Theta (t)\,, \label{4.1a} \ee where
$M=\exp\bigl(\sum_{i>0} m^{(i)}\bigr)$ and $ B \in G_0$, i.e.,  an
element of the zero  grade subgroup. It therefore follows that \be
\Theta {{\pa {\Theta^{-1}}}\o {\pa {t_{-n}}}} =  {{\pa }\o
{\pa_{t_{-n}}}}+ ( BM E^{(-n)}M^{-1}B^{-1})_{-}\,. \label{4.2} \ee
By construction $[\pa_{t_{-n}}, \pa_{t_m}]=0$ and henceforth \be
\biggl[\pa_x+E+A_0,\pa_{t_{-n}}+\sum_{i=1}^{n}D^{(-i)}\biggr]=
\Theta\left[\pa_x+ E,\pa{t_{-n}}\right]\Theta^{-1}=0\,.
\label{4.3} \ee Decomposing the zero curvature condition
(\ref{4.3}) into graded components we find \be\label{4.4}\ba{l}
 \pa_x D^{(-n)}_n+[A_0,D^{(-n)}_n]=0\,,\\[2pt]
 \pa_x D^{(-n+1)}_n[ E,D^{(-n)}_n]+[A_0,D^{(-n+1)}_n]=0\,,\\[2pt]
\qquad\vdots\\[2pt]
\pa_x D^{(-1)}_n[E,D^{(-2)}_n]+[A_0,D^{(-1)}_n]=0\,,\\[2pt]
\pa_{t_{-n}}A_0-[E,D^{(-1)}_n]=0\,.\ea\ee Eqns. (\ref{4.4}) can be
solved recursively, however notice that in general, contrary  to
$D^{(i)}$ in eqns. (\ref{2.3}), the $D^{(-i)}$  are nonlocal
functionals of the fields  $A_0$.  There is one particular case,
for $t= t_{-1}$, in which we obtain a closed local solution. Let
$n=1$ in eqn. (\ref{4.3}) \be
\left[\pa_x+E+A_0,\pa_{t_{-1}}+(BE^{(-1)}B^{-1})\right]=0\,.
\label{4.5} \ee If we now compare eqn. (\ref{4.5}) with
(\ref{zzc}) and (\ref{zz}) identifying $\bar z=-x$, $z=t_{-1}$ and
$E^{(\pm 1)}=\eps_{\pm}$ we find, \be
D^{(-1)}_1=B\eps_-B^{-1}\,,\quad A_0=\bar\pa BB^{-1}=-\pa_x
BB^{-1}\,. \label{4.6} \ee With  the space time identified as
above, it becomes clear that the Leznov--Saveliev eqns.
(\ref{1.1}) can be put within the general construction for
integrable hierarchies associated to negative grade time evolution
(\ref{4.3}).  This explains the relationship between  the
sinh--Gordon (\ref{1.2}) and the mKdV (\ref{2.4}) equations  as
well as, the Lund--Regge (\ref{1.4}) and the AKNS (\ref{2.5})
equations.

\section{Supersymmetric integrable hierarchies}

In this section we shall  consider how the structure of the Lax
operators changes when terms with  half--integer grades appear in
$\widehat \lie = \oplus_{n \in { Z}} \, \lie _{n/2}$. As a
consequence of such terms being present in the exponent of the
dressing matrices \be\label{4.1}\ba{rcl}
\Theta&=&\disty\exp\biggl(\sum_{i<0}\theta^{(i)}\biggr)=
\exp\left(\theta^{(-1/2)}+\theta^{(-1)}+\theta^{(-3/2)}+\ldots\right),\\[16pt]
M&=&\disty\exp\biggl(\sum_{i>0} m^{(i)}\biggr)=
\exp\left(m^{(1/2)}+m^{(1)}+m^{(3/2)}+\ldots\right)\ea\ee the form
of the Lax operator is changed as follows (compare with
(\ref{3.3})): \br \pa_{t_1} (\Theta) &=& (\Theta E
\Theta^{-1})_{-} \Theta = \big\lbrack \Theta E \Theta^{-1} -
(\Theta E \Theta^{-1})_{+}
\big\rbrack \Theta= \nonu \\
&= &\Theta  E + ( E+ \left[ \theta^{(-1)} , E \right] +
\left[ \theta^{(-1/2)} , E \right] + \h \left[ \theta^{(-1/2)} , \left[
\theta^{(-1/2)} , E \right]\right] ) \Theta= \nonu \\
&=&\Theta  E + ( E+ A_0 + A_{1/2} +k_0 ) \Theta\,. \label{3.1} \er
Here \br A_0 &=&\left[ \theta^{(-1)} , E \right] + \h \left[
\theta^{(-1/2)} , \left[
\theta^{(-1/2)} , E \right]\right] \Big\v_{\cal M}  \, \in \,{\cal M}\,, \label{azero}\\
A_{1/2}&=& \left[ \theta^{(-1/2)} , E \right] \, \in \,{\cal M}\,,
\label{aonehalf}\\
k_0 &=& \h \left[ \theta^{(-1/2)} , \left[ \theta^{(-1/2)} , E
\right]\right]\Big\v_{\cal K}\,\in \,{\cal K}\,, \label{kzero} \er
where $ \Big\v_{\cal K}$ and $\Big\v_{\cal M}$ denote projections
on the kernel ${\cal K}$ and image ${\cal M}$, respectively. This
shows that, in case of a half-integer grading, a general
expression for the Lax operator is \be {\cal L} = \pa_x + E+ A_0 +
A_{1/2} +k_0 \,. \label{laxoper} \ee The unconventional grade zero
term  $k_0$ residing in ${\cal K}$ appears here due to the
half--integer grading (encountered in case of $sl(2|1)$ with
principal gradation \cite{npb}).

Following the procedure explained in Sect. 2 with $\Theta$ given
in (\ref{4.1}) we generalize the zero curvature form of eqn.
(\ref{3.6}),i.e. \be\label{3.6a}\ba{c}
\disty\biggl[\pa_x+E+A_0+A_{1/2}+k_0,\pa_{t_n}+E^{(n)}+
\sum_{i=0}^{2n-1}D_n^{(i/2)}\biggr]=\\[12pt]
=\Theta\left[\pa_x+E,\pa_{t_n}+E^{(n)}\right]\Theta^{-1}=0\,,\ea\ee
which can be solved recursively for $D_n^{(i)}$ and
$D_n^{(i+1/2)}$, $i= 1,\cdots,n-1$. This also leads to the
equations of motion (time evolution ) for the fields $A_0$.

The negative grade sector (\ref{4.3}) can also be extended. From
Sect. 3 we also find \be\label{3.7}\ba{c}
\disty\biggl[\pa_x+E+A_0+A_{1/2}+k_0,\pa_{t_{-n}}+
\sum_{i=1}^{2n}D_n^{(-i/2)}\biggr]=\\[12pt]
=\Theta\left[\pa_x+E,\pa_{t_{-n}}\right]\Theta^{-1}=0\,.\ea\ee In
particular for $n=1$ we find \be
\left[\pa_x+E+A_0+A_{1/2}+k_0,\pa_{t_{-1}}+D_1^{(-1)}+D_1^{(-1/2)}\right]=0\,,
\label{3.8} \ee where from (\ref{4.2})
$$
D_1^{(-1)}=BE^{(-1)}B^{-1}\,,\quad
D_1^{(-1/2)}=Bj_{-1/2}B^{-1}\,,\quad
j_{-1/2}=\left[m^{(1/2)},E^{(-1)}\right].
$$
Taking the grade $-1$ of eqn. (\ref{3.8}) we find
$$
\pa_x(BE^{(-1)} B^{-1})+[A_0, BE^{(-1)}B^{-1}]=0\,,
$$
which has solution  $A_0 = -\pa_x BB^{-1}$. Take now the grade
$-{1\o 2}$ of eqn. (\ref{3.8}) to obtain respectively \be
\pa_x\left(j_{-1/2}\right)=[E^{(-1)},B^{-1}A_{1/2}B]\,.
\label{3.9} \ee From grades ${1\o 2}$ and zero, we obtain \br
\pa_{-1}\,(A_{1/2})&=&\left[E,B j_{-1/2}B^{-1}\right],\label{3.10}\\
\pa_{-1}(\pa_{x}B\,B^{-1})&=&\left[B E^{(-1)}B^{-1},E\right]+
\left[B j_{-1/2}B^{-1},A_{1/2}\right]. \label{3.11} \er Eqns.
(\ref{3.9})\,--\,(\ref{3.11}) correspond to the equations of
motion of the associated relativistic integrable model.

\section{The \bmth{N=2} super MKdV and sinh\bmth{-}Gordon equations}

As an application  consider the  loop algebra $\widehat {SL}(2,2)$
described in the appendix. In parametrizing the Lax components
$A_{1/2}$ and $A_0$  in (\ref{3.7}) we determine from
(\ref{aonehalf})  and (\ref{azero}) the first two terms of
$\Theta$ in (\ref{4.1}), namely $\theta_{-1/2}$ and $\theta_{-1}$.
The lowest grade $\theta_{-1/2}$, in turn, determines the quantity
$k_0$ from eqn. (\ref{kzero}). From \cite{npb} we found that the
existence of nontrivial $k_0$ gives rise to nonlocal supersymmetry
transformation.  In order to provide a simple example of local
$N=2$ supersymmetric integrable model we shall consider a
subalgebra of the loop algebra $\widehat {SL}(2,2)$ whose
generators are given by (\ref{sub}). Within such subalgebra we
make sure that $k_0=0$. Let  the Lax operator \be
L=\pa_x+E+A_0+A_{1/2} \label{laxoper1} \ee be specified by \be
E=K_1^{(1)}+K_2^{(1)}+I^{(1)},~~ A_0=u_1M_1^{(0)}+u_3M_3^{(0)},~~
A_{1/2}={\bar\psi}_1G_1^{(1/2)}+{\bar\psi}_3G_3^{(1/2)}.\ee We now
solve the zero--curvature equation (\ref{3.6}) for $n=3$. It is
explicitly given by \be\label{zc3}\ba{c}
\left[\pa_x+E+A_0+A_{1/2},\pa_3+\mathcal{D}+E^{(3)}\right]=0\,,\\[4pt]
\mathcal{D}=D_3^{(0)}+D_3^{(1/2)}+D_3^{(1)}+D_3^{(3/2)}+D_3^{(2)}+D_3^{(5/2)}\,,
\ea\ee where  $E^{(3)}= K_1^{(3)}+K_2^{(3)}+I^{(3)}$.  We then
obtain the following equations of motion for the $N=2$ super MKdV,
\br
4\pa_3\psi_1&=&\pa_x^3\psi_1-\sfrac{3}{2}\psi_1\pa_x(u_1^2+u_3^2)-
3\pa_x(\psi_1)(u_1^2+ u_3^2)-3\psi_3\pa_x(u_1u_3)\,,
\label{xi1eqsmot}\\
4\pa_3\psi_3&=&\pa_x^3\psi_3-\sfrac{3}{2}\psi_3\pa_x(u_1^2+u_3^2)-
3\pa_x(\psi_3)(u_1^2+ u_3^2)-3\psi_1\pa_x(u_1u_3)
\label{xi3eqsmot} \er and \br
\hskip-7mm
4\pa_3u_1&=&\pa_x\Bigl[\pa_x^2u_1-2u_1^3+
3u_1(\psi_1\pa_x\psi_1-\psi_3\pa_x\psi_3)+
3u_3(-\psi_1\pa_x\psi_3+\psi_3\pa_x\psi_1)\Bigr],\nonu \\
\hskip-7mm
4\pa_3u_3&=&\pa_x\Bigl[\pa_x^2u_3-2u_3^3-
3u_3(\psi_1\pa_x\psi_1-\psi_3\pa_x\psi_3)-
3u_1(-\psi_1\pa_x\psi_3+\psi_3\pa_x\psi_1)\Bigr].\label{bos} \er
These equations are invariant under the following supersymmetry
transformations \be
 \d u_1 = 2 \pa_x  (-\psi_1 \eps_2 +\psi_3 \eps_4  )\,,\quad
\d u_2 = 2 \pa_x (-\psi_1 \eps_4 +\psi_3 \eps_2 ) \ee and \be
\d\psi_1 = u_1 \eps_2 - u_3 \eps_4\,,\quad
\d\psi_3=u_1\eps_4-u_3\eps_2\,, \ee with $\eps_2$, $\eps_4$
constant grassmanian parameters.

For the relativistic case we parametrize $B=\E^{\p_1M_1+\p_3M_3}$
and \br A_0 &=& -{ \pa}_x B \, B^{-1} =  -\pa_x \phi_1 \, M_1 - {
\pa}_x \phi_3 M_3\,,
\label{aazero}\\
A_{1/2} &=& \psi_1 G_1^{(1/2)} + \psi_3 G_3^{(1/2)}\,,
\label{aahalf}\\
\jmath_{-1/2} &=& {\psi}_2 G_2^{(-1/2)} + {\psi}_4 G_4^{(-1/2)}\,,
\label{ajhalf}\\
E^{(-1)} &=& K_1^{(-1)}+K_2^{(-1)} +I^{(-1)}\,. \label{aeminus}\er
With notation \be {\bar\psi}_{\pm}=\psi_2\pm\psi_4\,,\quad
{\psi}_{\pm}=\psi_1\pm\psi_3\,,\quad \phi^{\pm}=\phi_1\pm\phi_3\ee
the eqns. of motion (\ref{3.9})\,--\,(\ref{3.11}) become \br
\pa_{-1}\psi_{\pm}&=&-2{\bar
\psi}_{\mp}\cosh\p_{\pm}\,,\label{n4p1chpsi}\\
\pa_x{\bar\psi}_{\pm}&=&-2\psi_{\mp}\cosh\p_{\pm}\,,\label{pabpsip}\\
\pa_{-1}\pa_x \phi_{+}&=&-4\sinh\phi_+\cosh\phi_-+
4\psi_+{\bar\psi}_+\sinh\phi_{-}\,,\label{n4eq5}\\
\pa_{-1}\pa_x\phi_{-}&=&-4\cosh\phi_+\sinh\phi_-+
4\psi_-{\bar\psi}_-\sinh\phi_+,.\label{n4eq6} \er These equations
are invariant under the supersymmetry transformations: \be \delta
\p_{\pm} = 2 \psi_{\mp} \eps_{\pm}\,,\quad \delta \psi_{\pm}= -
\pa_x \p_{\mp} \eps_{\pm}\,,\quad \delta {\bar \psi}_{\pm}= 2
\sinh \p_{\pm}  \eps_{\mp}\,, \label{sustrn2a} \ee where
$\eps^{\pm}=\eps_2\pm\eps_4$. The above $N=2$ Sine--Gordon
equations correspond to the ones proposed by Kobayashi and Uematsu
\cite{koba}  and written in the above form by Nepomechie
\cite{nepo}.  The construction of integrable models with higher
supersymmetries such as $N=4$, 8 involves higher rank subalgebras
of $\hat{SL}(4,4)$, $\hat{SL}(8,8)$~\cite{inprep}.

{\bf Acknowledgments}  
We thank Prof. G.M. Sotkov for numerous discussions and suggestions.
We are grateful to  CNPq and FAPESP  for 
financial support.

\section*{Appendix: Algebra \bmth{SL(2,2)}}

The super Lie algebra  $SL(2,2)$  is a  rank 3 algebra with simple
roots \be \a_1=e_1-e_2\,,\quad \a_2=e_2-f_1\,,\quad
\a_3=f_1-f_2,\quad e_i\cdot e_j=-f_i\cdot f_j=\d_{ij}\,.
\label{a.1} \ee The affine (loop) algebra $\widehat {SL}(2,2)$ is
given by \be\label{algebra}\ba{rcll}
\left[\tilde{h}_i^{(n)},E_{\a_j}^{(m)}\right]&=&(\a_i\cdot\a_j)E_{\a_j}^{(n+m)},\\[2pt]
\left[E_{\a_i}^{(n)},E_{-\a_i}^{(m)}\right]&=&
\mathrm{Str}\,(E_{\a_i},E_{-\a_i})\tilde{h}_i^{(n+m)},\quad&
i,\,j=1,\,2,\,3\,,\\[2pt]
\left[E_{\a}^{(n)},E_{\b}^{(m)}\right]&=&\eps(\a,\b)E_{\a+\b}^{(n+m)},\quad&
\a+\b\,\mbox{~is~root}\,,\\[2pt]
\left[E_{\a}^{(n)},E_{\b}^{(m)}\right]&=&0\,,\quad&
\mbox{otherwise}\,,\\[2pt]
\left[d,T^{(m)}\right]&=&m T^{(m)},\quad&
T^{(m)}=E_{\a}^{(m)}\,\mbox{or}\,h_i^{(m)}\,,\ea\ee where $n,m\in
Z$ or $n,m\in Z+{1\o 2}$ according to the bosonic or fermionic
character of the generator respectively. Define now
$$
\ba{rcl}
Q&=&d+\h(\tilde h_1^{(0)}+\tilde h_3^{(0)})\,,\\[6pt]
E&=&(E_{\a_1}^{(0)}+E_{-\a_1}^{(2)})+(E_{\a_3}^{(2)}+E_{-\a_3}^{(0)})+I^{(1)}\,.\ea
$$
The  fermionic and bosonic components of the Kernel ${\cal K} $
are generated by \be\label{a.2}\ba{rcl}
f_{1,\eta}^{(n+1/2)}&=&\disty\left(\eta E_{\a_1+\a_2}^{(n-1/2)}+
E_{-\a_1-\a_2}^{(n+3/2)}\right)+ \left(\eta
E_{\a_2+\a_3}^{(n+33/2)}+E_{-\a_2-\a_3}^{(n-1/2)}\right),\\[10pt]
f_{2,\eta}^{(n+1/2)}&=&\disty\left(\eta
E_{\a_1+\a_2+\a_3}^{(n+1/2)}+E_{-\a_1-\a_2-\a_3}^{(n+1/2)})+ (\eta
E_{\a_2}^{(n+1/2)}+E_{-\a_2}^{(n+1/2)}\right), \ea\ee $\eta=\pm1$
and \br K_1^{(n)}&=&E_{\a_1}^{(n-1)}+E_{-\a_1}^{(n+1)},\nonu\\
K_2^{(n)}&=&E_{\a_3}^{(n+1)} + E_{-\a_3}^{(n-1)},\label{a.3}\\
I^{(n)}&=&\tilde h_1^{(n)}+2\tilde h_2^{(n)}-\tilde h_3^{(n)}.
\nonu \er The image ${\cal M}$ by \be\label{aa.4}\ba{rcl}
g_{1,\eta}^{(n+1/2)}&=&\disty\left(\eta E_{\a_1+\a_2}^{(n-1/2)}+
E_{-\a_1-\a_2}^{(n+3/2)}\right)- \left(\eta
E_{\a_2+\a_3}^{(n+3/2)}+E_{-\a_2-\a_3}^{(n-1/2)}\right),\\[10pt]
g_{2,\eta}^{(n+1/2)}&=&\disty\left(\eta
E_{\a_1+\a_2+\a_3}^{(n+1/2)}+
E_{-\a_1-\a_2-\a_3}^{(n+1/2)}\right)- \left(\eta
E_{\a_2}^{(n+1/2)}+E_{-\a_2}^{(n+1/2)}\right), \ea\ee $\eta=\pm1$
and \be\label{a.4}\ba{rclrcl} M_1^{(n)}&=&\tilde{h}_1^{(n)},\quad&
M_2^{(n)}&=&-E_{\a_1}^{(n-1)}+E_{-\a_1}^{(n+1)},\\[3pt]
M_3^{(n)}&=&-{\tilde h}_3^{(n)},\quad&
M_4^{(n)}&=&-E_{\a_3}^{(n-1)}+E_{-\a_3}^{(n+1)}\ea\ee
respectively. Define now
$$
\ba{rclrcl}
F_1^{(n+1/2)}&=&\disty\frac{1}{\sqrt{2}}\left(f_{1,+}^{(n+1/2)}+
f_{2,+}^{(n+1/2)}\right),\quad&
F_2^{(n+1/2)}&=&\disty\frac{1}{\sqrt{2}}\left(f_{1,-}^{(n+1/2)}+
f_{2,-}^{(n+1/2)}\right),\\[10pt]
F_3^{(n+1/2)}&=&\disty\frac{1}{\sqrt{2}}\left(f_{1,+}^{(n+1/2)}-
f_{2,+}^{(n+1/2)}\right),\quad&
F_4^{(n+1/2)}&=&\disty\frac{1}{\sqrt{2}}\left(f_{1,-}^{(n+1/2)}-
f_{2,-}^{(n+1/2)}\right)\ea
$$
and
$$
\ba{rclrcl}
G_1^{(n+1/2)}&=&\disty\frac{1}{\sqrt{2}}\left(g_{1,+}^{(n+1/2)}+
g_{2,+}^{(n+1/2)}\right),\quad&
G_2^{(n+1/2)}&=&\disty\frac{1}{\sqrt{2}}\left(g_{1,-}^{(n+1/2)}+
g_{2,-}^{(n+1/2)}\right),\\[10pt]
G_3^{(n+1/2)}&=&\disty\frac{1}{\sqrt{2}}\left(g_{1,+}^{(n+1/2)}-
g_{2,+}^{(n+1/2)}\right),\quad&
G_4^{(n+1/2)}&=&\disty\frac{1}{\sqrt{2}}\left(g_{1,-}^{(n+1/2)}-
g_{2,-}^{(n+1/2)}\right).\ea
$$
We now define a consistent subalgebra of the affine
$\widehat{SL}(2,2)$ loop algebra by selecting the following
generators, \be\ba{l} M_1^{(2n)},~~M_3^{(2n)},~~ M_2^{(2n+1)},~~
M_4^{(2n+1)},~~ K_1^{(2n+1)},~~ K_2^{(2n+1)},~~I^{(2n+1)},\\[4pt]
G_1^{(2n+1/2)},~~ G_3^{(2n+1/2)},~~ F_2^{(2n+1/2)},~~
F_4^{(2n+1/2)},\\[4pt]
G_2^{(2n+3/2)},~~ G_4^{(2n+3/2)},~~ F_1^{(2n+3/2)},~~
F_3^{(2n+3/2)}\hskip20mm\mbox{for $n\in Z$.} \ea \label{sub} \ee

\end {document}